\documentclass[twocolumn,prb]{revtex4}
\usepackage{amsfonts}
\usepackage{amsmath}
\usepackage{amssymb}
\usepackage{graphicx}

\begin{document}

\preprint{APS/123-QED}
\title{Coherent sliding dynamics and spin motive force driven by crossed
magnetic fields in chiral helimagnet}
\author{Jun-ichiro Kishine}
\affiliation{Division of Natural and Environmental Sciences, The Open University of
Japan, Chiba, 261-8586, Japan}
\author{I. G. Bostrem}
\affiliation{Institute of Natural Sciences, Ural Federal University, Ekaterinburg,
620083, Russia}
\author{A. S. Ovchinnikov}
\affiliation{Institute of Natural Sciences, Ural Federal University, Ekaterinburg,
620083, Russia}
\author{Vl. E. Sinitsyn}
\affiliation{Institute of Natural Sciences, Ural Federal University, Ekaterinburg,
620083, Russia}
\date{\today }

\begin{abstract}
We demonstrated that the chiral soliton lattice formed out of a chiral
helimagnet exhibits coherent sliding motion by applying a time-dependent
magnetic field parallel to the helical axis, in addition to a static field
perpendicular to the helical axis. To describe the coherent sliding, we use
the collective coordinate method and numerical analysis. We also show that
the time-dependent sliding velocity causes a time-varying Berry cap which
causes the spin-motive-force. A salient feature of the chiral soliton
lattice is appearance of the strongly amplified spin motive force which is
directly proportional to the macroscopic number of solitons (magnetic kinks).
\end{abstract}

\pacs{Valid PACS appear here}
\maketitle

\section{Introduction}

Spin-based-electronics (spintronics) is now an emerging field. An essential
notion behind this emergence is the fact that the `spin magnetic current'
without relying on electric current would magnificently reduce the energy
loss and switching time during the information read/write processes. At the
heart of spintronics is to drive motion of magnetic textures in a
controllable manner. There are two ways to drive the motion, i.e.,
incoherent and coherent methods.\cite{Sonin2010} The incoherent method is
typically realized by injecting a spin-polarized current into a sample.\cite%
{Zutic04} On the other hand, the coherent method is realized in a
magnetically ordered state by twisting the phase angle of the magnetic order
parameter which directly couples to a magnetic field. In the coherent
method, the phase rigidity (stiffness) of the whole spin system makes it
possible to transmit the spin rotation at one end of a sample to the other
end via spin torque transfer. Although the coherent method has potential
advantage, it has not been extensively studied in the context of the
present-day spintronics, because it is not so easy to prepare rigid phase
object which transports experimentally measured quantity in a controllable
manner. To seek for the possibility of the coherent method is interesting.

In this paper we propose that chiral helimagnets are promising candidates to
realize the coherent method. The chiral helimagnetic (CHM) state is
characterized by the vector spin chirality as an order parameter. The
structure is stabilized by the antisymmetric Dzyaloshinskii-Moriya (DM)
interaction and realized in crystals without rotoinversion symmetry. A
guiding principle for materializing this effect is symmetry-adapted material
synthesis, i.e., the interplay of crystallographic and magnetic chirality
plays a key role there.

\begin{figure}[b]
\begin{center}
\includegraphics[width=75mm]{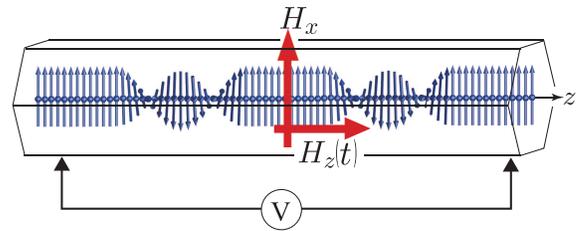}
\end{center}
\caption{Schematic picture of the CSL. SMF generation needs static
transverse field $H_{x}$ and time-dependent longitudinal field $H_{x}$.}
\label{CSL}
\end{figure}

The CHM state is, however, nothing more than non-colinear linear (harmonic)
spin structure. To carry physically measurable quantity, we need nonlinear
structure. Fortunately, under the magnetic field applied perpendicular to
the helical axis, the CHM is transformed to a non-linear magnetic structure
called a chiral soliton lattice (CSL) [see Fig.~\ref{CSL}] which is
equivalent to a magnetic kink crystal (MKC).\cite{Dzyaloshinskii64,KIY05} In
the CSL state, the ground state possesses a periodic array of the
commensurate (C) and incommensurate (IC) domains partitioned by
discommensurations. Recently, using Lorenz microscopy and small-angle
electron diffraction, the CSL was experimentally verified in a hexagonal
chiral magnet Cr$_{1/3}$NbS$_{2}$.\cite{Togawa2012} Present authors have
discussed physical outcome of the CSL state from various viewpoints. \cite%
{BKO2008,KO2009,KOP2010,KOP2011}

As pointed out in Refs.\cite{BKO2008,KOP2010}, the CSL exhibits coherent
collective sliding motion in non-equilibrium state. Once the sliding is
triggered, the CSL maintains its persistent motion assisted by the dynamical
generation of the inertial mass. The mass generation is understood by D\"{o}%
ring-Becker-Kittel mechanism of the moving domain wall.\cite%
{Doring48,Becker50,Kittel50} In this mechanism, the longitudinal
(out-of-plane) component of the slanted magnetic moment inside the domain
wall emerges as a consequence of translational motion. An additional
magnetic energy associated with the resultant demagnetization field is
interpreted as the kinetic energy of the wall.

The incoherent current injection method to drive the sliding was already
proposed by the present authors.\cite{KOP2011} In this paper, we demonstrate
that crossed magnetic fields are eligible to cause the coherent motion of
the whole CSL. Here, we mean by the crossed fields that in addition to a
static field perpendicular to the helical axis, which stabilizes the CSL\
formation, a magnetic field parallel to the helical axis is imposed.

Once the coherent motion occurs, the natural question arises as to whether
the motion has observable consequences for the spin motive force (SMF).\cite%
{Barnes2007} Quite naturally to expect that the time dependence of the
longitudinal magnetic field manifests itself in a temporal regime of the
SMF. Time dependences of the spinmotive forces are classified into three
types, i.e., (i) transient, (ii) continous ac, (iii) and continous dc ones.
For example, the domain wall motion\cite{Yang2009} and electron transport
through ferromagnetic nanoparticles\cite{Hai2009} lead to the type (i) SMF.
A vortex core dynamics of a magnetic disk caused by an oscillating magnetic
field directed in the disk plane induces a continous ac spinmotive force of
type (ii).\cite{Ohe2009} A resonant microwave excitation of a comb-shaped
ferromagnetic thin film produces a continous dc spinmotive force of type
(iii).\cite{Yamane2011} We will demonstrate that the time-dependent
longitudinal field, as shown in Fig.~\ref{CSL}, possibly causes the SMF of
type (i) and (ii) in the chiral helimagnet. As a remarkable feature, we note
that our CSL is a macroscopically ordered object, which contains macroscopic
amounts of magnetic solitons. Due to this huge number of solitons, the SMF
is expected to be strongly amplified as compared with the SMF caused by a
single magnetic domain wall in a ferromagnet.

In Sec. I, we present a model and summarize necessary background on the CSL
dynamics. In Sec. II, we demonstrate that under the presence of the
longitudinal field, the CSL becomes unstable and coherent motion occurs. In
Sec. III, we present numerical analysis of dynamics to support analytical
consideration presented in Sec. III. In Sec. IV, we discuss the SMF
associated with the coherent motion. We conclude our results in Sec. V.

\section{Static deformation of CSL under \textbf{crossed magnetic fields}}

\subsection{Basic equations of chiral soliton lattice}

\subsubsection{Static structure}

Mono-axial chiral helimagnet is described by an effective one-dimensional
Hamiltonian,%
\begin{equation}
H=-J\sum_{i}\boldsymbol{S}_{i}\cdot\boldsymbol{S}_{i+1}-\boldsymbol{D}%
\cdot\sum_{i}\boldsymbol{S}_{i}\times\boldsymbol{S}_{i+1}+\boldsymbol{%
\tilde {H}}\cdot\sum_{i}\boldsymbol{S}_{i},  \label{lattH}
\end{equation}
where $\mathbf{S}_{i}$ is the local spin moment at the site $i$, $J>0$ is
the nearest-neighbor ferromagnetic exchange interaction, $\mathbf{D}=D\hat{%
\mathbf{e}}_{z}$ is the mono-axial Dzyaloshinskii-Moriya (DM) interaction
along a certain crystallographic chiral axis (taken as the $z$-axis). We
take $z$-axis as the mono-axis and apply magnetic field $\boldsymbol{\tilde{H%
}}=g\mu_{B}\boldsymbol{H}=g\mu_{B}(H_{x},0,H_{z})$ in the $xz$-plane, where $%
g$\ is the electron g-factor and $\mu_{B}=\left\vert e\right\vert \hbar/2m$
is the Bohr magneton.

In the semi-classical approach, because of the slowly varying nature of the
spin variables, it is legitimate to introduce the continuous field variable $%
\boldsymbol{S}(z)=\sum\nolimits_{i}\boldsymbol{S}_{i}\delta(z-z_{i}) \equiv S%
\boldsymbol{n}(z) $. A{\ unit vector field }$\boldsymbol{n}(z)${\ is
represented as }%
\begin{equation}
\boldsymbol{n}(z)={[\sin\theta{(z)}\cos\varphi{(z)},\sin\theta{(z)}\sin
\varphi{(z)},\cos\theta{(z)}],}  \label{Polar}
\end{equation}
by using the polar angles $\theta{(z)}${\ and }$\varphi{(z)}$. The continuum
version of the Hamiltonian (\ref{lattH}), $H=\int\nolimits_{0}^{L}dz\mathcal{%
H}$, where $L$\ denotes the linear dimension of the system, includes the
Hamiltonian density, 
\begin{equation}
\mathcal{H}=\frac{JS^{2}}{2a_{0}}\left( \partial_{z}\boldsymbol{n}\right)
^{2}-\frac{S^{2}}{a_{0}^{2}}\text{ }\boldsymbol{D}\cdot\boldsymbol{n}%
\times\partial_{z}\boldsymbol{n}+\frac{S}{a_{0}^{3}}\boldsymbol{\tilde{H}}%
\cdot\boldsymbol{n}.  \label{contH}
\end{equation}
Here $a_{0}$\ is the atomic lattice constant along the chiral axis ($%
a_{0}\simeq10$\AA\ in Cr${}_{1/3}$NbS$_{2}$.\cite{Togawa2012})

For $\boldsymbol{H}=0$, the Hamiltonian (\ref{lattH}) gives a $xy$-planer
helimagnetic structure, $\theta{(z)}=\pi/2$ and ${\varphi{(z)=Q}}_{0}z$ with
the modulation wave number being given by $Q_{0}=a_{0}^{-1}\arctan(D/J)%
\simeq a_{0}^{-1}D/J$.

For only non-zero transverse field perpendicular to the chiral axis, the CSL
structure becomes the ground state characterized by $\theta=\pi/2$ and%
\begin{equation*}
\varphi_{0}(z)=2\mathrm{am}\left( \dfrac{\pi Q_{0}}{4E}z\right) ,
\end{equation*}
where $\mathrm{am}$ is the Jacobi's amplitude function with the elliptic
modulus $\kappa$ ($0\leq\kappa<1$), and $E=E(\kappa)$ is the complete
elliptic integral of the second kind.\ The elliptic modulus $\kappa$ is
determined from the condition, 
\begin{equation}
\kappa=\frac{4E}{\pi Q_{0}a_{0}}\sqrt{\frac{\tilde{H}_{x}}{JS}}.
\label{kappa}
\end{equation}
This equation is also written as $\tilde{H}_{x}/\tilde{H}{_{c}}=\left(
\kappa/E\right) ^{2}$ by introducing the critical field corresponding to $%
\kappa=1$, 
\begin{equation}
\tilde{H}_{c}=(\pi Q_{0}/4)^{2}JSa_{0}^{2}\sim D^{2}/J,  \label{criticalH}
\end{equation}
at which a incommensurate-to-commensurate phase transition occurs. The
spatial period of the CSL is given by $L_{\text{CSL}}=8{KE/\pi Q}_{0}{,}$%
which continuously increases from $L{{_{\text{CHM}}=}}2\pi{/Q}_{0}$\ to
infinity when the magnetic field increases from zero to ${H_{c}}$. Here, $L{{%
_{\text{CHM}}}}\ $is the spatial period of CHM under zero field. $%
K=K(\kappa) $ is the complete elliptic integral of the first kind.

\subsubsection{Individual spin dynamics}

Next, we write down the basic equations for dynamics. Using the Hamiltonian
density (\ref{contH}), we construct the Lagrangian density,%
\begin{equation}
\tilde{\mathcal{L}}=\frac{\hbar S}{a_{0}^{3}}\left( \cos\theta-1\right)
\,\partial_{t}\varphi-\mathcal{H}.
\end{equation}
To incorporate the damping effect, we use the Rayleigh dissipation described
by 
\begin{equation}
\tilde{\mathcal{W}}=\frac{\alpha\hbar S}{2a_{0}^{3}}\left( \partial _{t}%
\boldsymbol{n}\right) ^{2}  \label{Gilbert}
\end{equation}
with $\alpha$ being a small coefficient specifying the Gilbert damping. The
Euler-Lagrange equations of motion is then given by 
\begin{subequations}
\begin{align}
\frac{\hbar S}{a_{0}^{3}}\sin\theta\,\partial_{t}\theta & =\frac {\delta%
\mathcal{H}}{\delta\varphi}+\frac{\delta\tilde{\mathcal{W}}}{\delta \dot{%
\varphi}},\text{ }  \label{ELEOMtheta} \\
\frac{\hbar S}{a_{0}^{3}}\sin\theta\,\partial_{t}\varphi & =-\frac {\delta%
\mathcal{H}}{\delta\theta}-\frac{\delta\tilde{\mathcal{W}}}{\delta \dot{%
\theta}},  \label{ELEOMphi}
\end{align}
which leads to the Landau-Lifshitz-Gilbert (LLG) equations, 
\end{subequations}
\begin{subequations}
\begin{align}
\hbar S\sin\theta\partial_{t}\theta & =-JS^{2}a_{0}^{2}\left\{ \sin ^{2}{%
\theta\partial_{z}^{2}{\varphi+}}\sin2{\theta}\left( \partial _{z}{\varphi}%
\right) \left( \partial_{z}{\theta}\right) \right\}  \notag \\
& +DS^{2}a_{0}\sin2{\theta}\left( \partial_{z}{\theta}\right) {-}\tilde {H}%
_{x}S{\sin\theta\sin\varphi}  \notag \\
& +\alpha\hbar S\sin^{2}{\theta}\left( \partial_{t}{\varphi}\right) ,
\label{EOMtheta} \\
\hbar S\sin\theta\partial_{t}\varphi & =JS^{2}a_{0}^{2}\left\{ \partial
_{z}^{2}{\theta-}\tfrac{1}{2}\sin2{\theta\left( \partial_{z}{\varphi}\right)
^{2}}\right\}  \notag \\
& +DS^{2}a_{0}\sin2{\theta}\left( {\partial_{z}{\varphi}}\right) -\tilde {H}%
_{x}S{\cos\theta\cos\varphi}  \notag \\
& +\tilde{H}_{z}S\sin{\theta}-\alpha\hbar S\partial_{t}{\theta}.
\label{EOMphi}
\end{align}
The LGG equations describe the individual (not collective) spin dynamics.

\subsubsection{Collective dynamics}

To consider the sliding motion of the CSL, we use the collective coordinate
method\cite{Christ-Lee75} used introduced in Ref.\cite{KOP2010}. {In this
formulation, the CSL dynamics is fully described by two dynamical variables,
the center of mass position, ${Z}$, and the }out-of-plane quasi-zero mode
(OPQZ) coordinate $\xi_{0}$. Using them the sliding solution is written as 
\end{subequations}
\begin{subequations}
\begin{align}
\varphi(z,t) & =\varphi_{0}\left[ z-Z(t)\right] ,  \label{collective_phi} \\
\theta(z,t) & =\pi/2+\xi_{0}(t)u_{0}\left[ z-Z(t)\right] .
\label{collective_theta}
\end{align}
The zero-mode wave function 
\end{subequations}
\begin{align}
{u}_{0}(z) & =\sqrt{\frac{K}{LE}}\mathrm{dn\,}\left( \dfrac{\pi Q_{0}}{4E}%
z\right) =\frac{2}{\pi Q_{0}}\sqrt{\frac{KE}{L}}\partial_{z}\varphi _{0}(z),
\label{zero_mode}
\end{align}
serves as the basis function of the $\theta$-fluctuations localized around
each soliton and $\xi_{0}(t)$\ is the OPQZ \textit{coordinate}. Here $%
\mathrm{dn\,}$ is the Jacobi-dn function. The function ${u}_{0}(z)$\ exactly
corresponds to the topological charge distribution because $\partial
_{z}\varphi_{0}(z)$. Using these variables, the Lagrangian, $\mathcal{L}%
=\int dz\tilde{\mathcal{L}}dz$, and the Rayleigh term, $\mathcal{W}=\int dz%
\tilde{\mathcal{W}}$, are respectively written as%
\begin{equation}
\mathcal{L}=\frac{\hbar S}{a_{0}^{3}}\mathcal{K}\xi_{0}\dot{Z}-\frac {%
\varepsilon_{0}^{(\theta)}}{a_{0}^{3}}\xi_{0}^{2}+\frac{S\sqrt{L}}{a_{0}^{3}}%
\tilde{H}_{z}\xi_{0},  \label{Collective Lagrangian}
\end{equation}
and 
\begin{equation}
\mathcal{W}=\frac{\alpha\hbar S}{2a_{0}^{3}}\left( \mathcal{M}\dot{Z}^{2}+%
\dot{\xi}_{0}^{2}\right) ,
\end{equation}
where $\mathcal{K}=\int\nolimits_{0}^{L}dzu_{0}(z)\partial_{z}\varphi_{0}(z)$
and $\mathcal{K}=\int\nolimits_{0}^{L}dz\left( \partial_{z}\varphi
_{0}(z)\right) ^{2}$ as given in Ref.\cite{KOP2010}. Furthermore, $%
\varepsilon_{0}^{(\theta)}\simeq D^{2}S^{2}/2J$ is an energy gap of the $%
\theta$-mode caused by the DM interaction which plays a role of easy-plane
anisotropy. We here note a useful relation%
\begin{equation}
\varepsilon_{0}^{(\theta)}/\tilde{H}_{c}=8S/\pi^{2},  \label{gap}
\end{equation}
[see Eq. (16) of Ref.\cite{KOP2010}]

Under the condition of weak field, $\kappa\ll1$, we have $\mathcal{K}\simeq
Q_{0}\sqrt{L}$ and $\mathcal{M}\simeq Q_{0}^{2}L$. We have $\xi_{0}(t)\neq0$
only for nonequilbrium state where the CSL exhibits sliding motion. An
emergence of such coherent collective transport in non-equilibrium state is
a manifestation of the dynamical off-diagonal long range order. Using Eqs. (%
\ref{collective_phi}) and (\ref{collective_theta}), Eqs. (\ref{EOMtheta})
and (\ref{EOMphi}) lead to equations of motion for collective dynamics, 
\begin{subequations}
\begin{align}
\hbar\mathcal{K}\dot{\xi}_{0} & =-\alpha\hbar\mathcal{M}\dot{Z},
\label{CEOM1} \\
\hbar\mathcal{K}\dot{Z} & =2{S^{-1}}\varepsilon_{0}^{(\theta)}\xi_{0}+\alpha%
\hbar\dot{\xi}_{0}-\sqrt{L}\tilde{H}_{z}.  \label{CEOM2}
\end{align}
This set of EOMs differs from Eqs. (37) given in Ref.\cite{KOP2010} in that
the sd interaction is absent and the longitudinal field is present. In Ref.%
\cite{KOP2010}, the incoherent driving of the CSL sliding motion was driven
by the spin-torque-transfer from the spin-polarized current to the local
spins. On the other hand, in the present case, we are discussing the
coherent driving caused by the uniform time-dependent magnetic field $\tilde{%
H}_{z}$.

Eqs. (\ref{CEOM1}) and (\ref{CEOM2}) are readily solved to give 
\end{subequations}
\begin{subequations}
\begin{align}
\dot{Z}(t) & =Ce^{-t/\tau_{\text{CSL}}}  \notag \\
& -\frac{e^{-t/\tau_{\text{CSL}}}}{(1+\alpha^{2})\hbar Q_{0}}\int
^{t}e^{t^{\prime}/\tau_{\text{CSL}}}\frac{d{\tilde{H}}_{z}(t^{\prime})}{%
dt^{\prime}}dt^{\prime},  \label{generalformulaforV} \\
\xi_{0}(t) & =De^{-t/\tau_{\text{CSL}}}  \notag \\
& +\frac{\alpha\sqrt{L}e^{-t/\tau_{\text{CSL}}}}{(1+\alpha^{2})\hbar}\int
^{t}e^{t^{\prime}/\tau_{\text{CSL}}}\tilde{H}_{z}(t^{\prime})dt^{\prime},
\label{generalformulaforxi}
\end{align}
where we used the relation $\mathcal{M}/\mathcal{K}^{2}=1$ for a weak field.
Constants $C$ and $D$ are determined by the initial condition $\dot{Z}(0)=0$
and $\xi_{0}(0)=0$. The intrinsic relaxation time of the CSL, caused by the
Gilbert damping, is introduced by 
\end{subequations}
\begin{equation}
\tau_{\text{CSL}}=\frac{\hbar S(\alpha+\alpha^{-1})}{2\varepsilon_{0}^{(%
\theta)}}\simeq\frac{\hbar(\alpha+\alpha^{-1})}{S}\frac{J}{D^{2}},
\label{Relaxation time}
\end{equation}
which is also written as 
\begin{equation}
1/\tau_{\text{CSL}}\simeq\alpha\omega_{\text{gap}},  \label{taugap}
\end{equation}
with $\omega_{\text{gap}}=\varepsilon_{0}^{(\theta)}/\hbar$ being a
characteristic frequency of the gap. In the case of static $\tilde{H}_{z}$,
we have trivial relaxational dynamics where the sliding motion never
persists. Furthermore, the DM\ interaction $D$ gives rise to a finite
relaxation time. Eq. (\ref{generalformulaforxi}) means that the longitudinal
field first directly coupes to $\xi_{0}$ and drives its grows via the
Gilbert damping process. Then, the sliding motion follows the growth of $%
\xi_{0}$. This process is consistent with intuitive ideas developed by D\"{o}%
ring.\cite{Doring48}

We here emphasize that the two coordinates $\xi_{0}$\ and $Z$\ coupled to
each other via the Gilbert damping $\alpha$ [see Eq. (\ref{CEOM1})]. If
there were no damping, we would have no correlated dynamics. This fact means
that the CSL\ never realizes dissipationless collective motion. As we will
see in Sec. V and appendix C, the presence of the damping is essential to
drive SMF.

\subsubsection{Comparison of material parameters to theoretical formulae}

It may be worthwhile to summarize relation between experimental data and
theoretical parameters by taking an example of Cr${}_{1/3}$NbS$_{2}$\cite%
{Togawa2012} and theoretical formulae. In this sample, it is reported that $%
a_{0}=1.212\times10^{-9}$m, $L{{_{\text{CHM}}}}=2\pi/Q_{0}=4.8\times10^{-8}$%
m. Cr atoms are in the trivalent state and have localized electrons with
spins of $S=3/2$. An observed critical field is $H_{\text{c}}=2300$Oe
corresponding to 0.31K. We have an estimation for the ratio, $%
D/J=\tan(Q_{0}a_{0})=0.16$. Another important quantity is the excitation gap
which is estimated as $\varepsilon_{0}^{(\theta)}\simeq$0.38K by using Eq. (%
\ref{gap}). The intrinsic relaxation time of the CSL is also estimated as $%
\tau_{\text{CSL}}\simeq(\alpha+\alpha^{-1})\times3.0\times10^{-11}$sec. A
small damping such as $\alpha\simeq10^{-2}$ leads to $\tau_{\text{CSL}%
}\sim3.0\times10^{-9}$sec. Smaller damping causes longer relaxation time. To
realize a longer period of relaxation processes, it is desirable to realize
a smaller value of $\alpha$ and a smaller gap frequency $\omega_{\text{gap}}$%
.

\subsection{Static deformation of CSL}

Now that we have set up all the necessary equations for dynamics, we proceed
with the stability analysis of the CSL\ against crossed magnetic fields.
Before going to dynamical deformation, it is worth while to study the static
deformation. For analysis, we here consider the weak field limit, $%
\left\vert \tilde{H}_{x}\right\vert \ll JS$ and $\left\vert \tilde{H}_{z}/%
\tilde{H}_{x}\right\vert \ll1$.

\begin{figure}[t]
\begin{center}
\includegraphics[width=50mm]{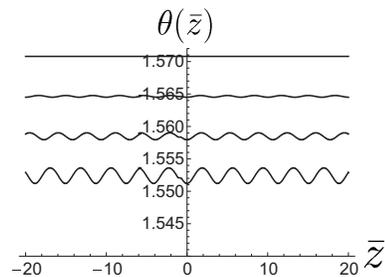}
\end{center}
\caption{Spatial modulation of $\protect\theta({\bar{z}})=\protect\pi/2+%
\protect\theta_{1}({\bar{z}})$ for $\tilde{H}_{x}/\tilde{H}%
_{c}=0,\,0.1,\,0.2,\,0.3$ (from above to below) with keeping $\tilde{H}_{z}/%
\tilde{H}_{x}=0.1$. Here a dimensionless coordinate $\bar{z}=\protect\pi %
Q_{0} z/4E$ is used.}
\label{TetReal}
\end{figure}

We introduce the static deformations as $\theta(z)=\pi/2+s\tilde{\theta}(z)$%
\ and $\varphi(z)=\varphi_{0}(z)+s\tilde{\varphi}(z)$, with the small
fluctuations $\tilde{\theta}$, $\tilde{\varphi}$ and $s$ being a dummy
expansion parameter. As derived in appendix A, we have 
\begin{equation}
\tilde{\theta}({\bar{z}})=\dfrac{\tilde{H}_{z}}{\tilde{H}_{x}}\dfrac {%
\kappa^{2}}{W}\left( C_{1}\tilde{\varphi}_{1}({\bar{z}})+C_{2}\tilde {\varphi%
}_{2}({\bar{z}})\right) ,  \label{EllipticTheta}
\end{equation}
where the Wronskian $W$ is given in Eq. (\ref{WronLame}) and the
coefficients $C_{1,2}$ are given by Eqs. (\ref{FTC1}) and (\ref{FTC2}). The
functions $\tilde{\varphi}_{1,2}$ are a pair of the fundamental solutions of
Lam\'e equation [Eq. (\ref{Fundament})]. In Fig. \ref{TetReal} we show the
obtained spatial modulation of $\theta({\bar{z}})=\pi/2+\theta_{1}({\bar{z}}%
) $ for various $\tilde{H}_{z}/\tilde{H}_{c}$ with keeping $\tilde{H}_{x}/%
\tilde {H}_{z}=0.1$. Since the coefficients $c_{1,2\,n}$ fall exponentially
with growth of $n$, we retain only terms with $n=\pm1$ which dominate the
terms with $\left\vert n\right\vert \geq2$. We see that finite $\tilde{H}%
_{z} $ tends to orient the spins toward the $z$-direction but causes \textit{%
non-uniform} spatial oscillation. This oscillation implies the static
defomation considered here is unstable against dynamic deformation. We will
see this dynamic deformation corresponds to the sliding motion of the CSL. 

\section{Coherent sliding dynamics by crossed magnetic fields}

\subsection{Energy and momeutum associated with the sliding motion}

In this section we will show that the static deformation of the CSL is
unstable against the dynamical instability, i.e., spontaneous coherent
sliding motion of the whole CSL. As the present authors previously pointed
out,\cite{BKLO2009} the Lagrangian constructed from the Hamiltonian Eq. (\ref%
{contH}) has the hidden Galilean symmetry induced thorugh Lie analysis.

This hidden symmetry justfies that the CSL has a linear momentum, 
\begin{equation}
P_{z}=\frac{\hbar S}{a_{0}^{3}}\int dz\left( 1-\cos\theta\right)
\partial_{z}\varphi,  \label{spin momentum}
\end{equation}
associated with the kinematic Berry phase. We connect the momentum variation 
$\delta P_{z}$ to the energy variation,\cite{Kosevich1990}%
\begin{equation}
\delta E=\int dz\left( \frac{\partial\mathcal{H}}{\partial\theta}\delta
\theta+\frac{\partial\mathcal{H}}{\partial\varphi}\delta\varphi\right) ,
\label{Variation of E}
\end{equation}
associated with the sliding motion. In the absence of the dissipation,
plugging the EOMs (\ref{ELEOMtheta}) and (\ref{ELEOMphi}) into Eq. (\ref%
{Variation of E}), we have%
\begin{equation}
\delta E=\hbar S\int dz\sin\theta\left( \frac{\partial\theta}{\partial t}%
\delta\varphi-\frac{\partial\varphi}{\partial t}\delta\theta\right) .
\label{Energy}
\end{equation}
The sliding motion means $\theta$\ and $\varphi\ $are funcitons of $u=z-Z(t)$
and we can make the replacement, $\partial_{t}\rightarrow-\dot{Z}%
\partial_{u} $, $\partial_{z}\rightarrow\partial_{u}$. Then, we easily
obtain the relation,%
\begin{equation}
\delta E=\dot{Z}\delta P_{z}.  \label{Goal}
\end{equation}
Based on this relation, we see that the coherent sliding occurs if the
condition $\dot{Z}\delta P_{z}<0$ is sataified for a given momentum transfer 
$\delta P_{z}$\ from the environment. We will see that in the present case
the longitudinal magnetic field $H_{z}$\ gives the momentum transfer and
drives the sliding motion.

\subsection{Coherent sliding caused by a transient longitudinal field}

We first consider a transient longitudinal field%
\begin{equation}
H_{z}(t)=H_{z0}(1-e^{-t/T}),  \label{tdB}
\end{equation}%
switched on in addition to a perpendicular static field $H_{x}$ which
stabilizes the CSL. Eq. (\ref{generalformulaforV}) gives the sliding
velocity,%
\begin{equation}
\dot{Z}(t)=-V_{0}\frac{\tau _{\text{CSL}}}{\tau _{\text{CSL}}-T}\left(
e^{-t/\tau _{\text{CSL}}}-e^{-t/T}\right) ,  \label{Velocity}
\end{equation}%
where the characteristic velocity is defined by 
\begin{equation}
V_{0}=\frac{\tilde{H}_{z0}}{\hbar Q_{0}(1+\alpha ^{2})}.
\end{equation}%
Eq. (\ref{Velocity}) indicates $\dot{Z}<0$ for $T<\tau _{\text{CSL}}$ and
then the condition $\dot{Z}\delta P_{z}<0$\ is satisfied. It is to be noted
that if the chirality of a crystal were inverted, i.e., $Q_{0}$ is inverted
to $-Q_{0}$, the velocity would be inverted. So, \textit{the sliding
orientation and the crystal chirality correlate to each other.}

In the case of Cr$_{1/3}$NbS$_{2}$, The characteristic velocity is estimated
as $V_{0}\simeq$0.13[m$\cdot$s$^{-1}\cdot$Oe$^{-1}$]. So, the sudden
switching of the longitudinal magnetic field $H_{z}$, satisfing the
condition $T<\tau_{\text{CSL}}$,\ will easily cause the coherent sliding
motion the CSL.

\begin{figure}[t]
\begin{center}
\includegraphics[width=50mm]{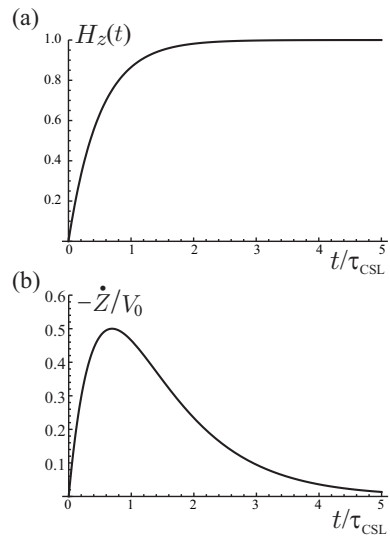}
\end{center}
\caption{Time dependence of (a) longitudinal field $%
H_{z}(t)=H_{z0}(1-e^{-t/T})$ $\ $and (b) velocity $\dot{Z}/V_{0}$ for $%
T=0.5\,\protect\tau _{\text{CSL}}$.}
\label{Relax}
\end{figure}

In Fig. \ref{Relax}, we show time-evolution of the sliding velocity. We see
the velocity grows linearly with time shortly after the field $H_{z}$\ is
switching on. Then, after the relaxation time of the filed, $T$, the
velocity begins to relax. Then, it finally relaxes to zero over the time
scale of the Gilbert damping $\tau_{\text{CSL}}$. Therefore, to realize
longer-lasting sliding motion, a smaller value of $\alpha$ and a smaller gap
frequency $\omega_{\text{gap}}$ may be desirable.

\subsection{Coherent oscillating motion under AC field}

\begin{figure}[t]
\begin{center}
\includegraphics[width=60mm]{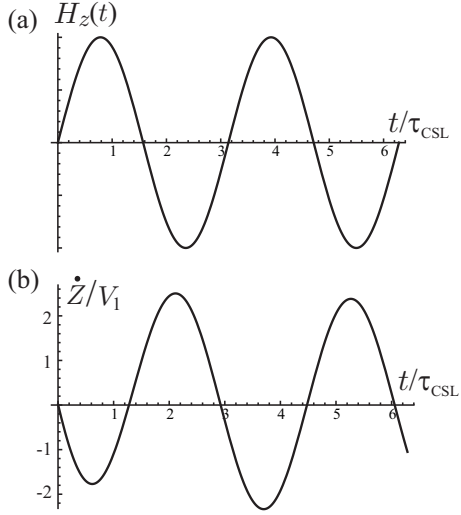}
\end{center}
\caption{Time dependence of (a) longitudinal field $H_{z}(t)=H_{z1}\sin(%
\Omega t)$ $\ $and (b) velocity $\dot{Z}/V_{0}$ for $\Omega^{-1}=0.5\,%
\protect\tau _{\text{CSL}}$.}
\label{ACCSL}
\end{figure}

It is also possible that an oscillating longitudinal field 
\begin{equation}
H_{z}(t)=H_{z1}\sin(\Omega t),
\end{equation}
causes a coherent oscillating motion of the CSL in addition to a
perpendicular static field $H_{x}$. In this case, Eq. Eq. (\ref%
{generalformulaforV})\ gives the velocity,%
\begin{equation}
\dot{Z}(t)=V_{1}[e^{-t/\tau_{\text{CSL}}}-\Omega\tau_{\text{CSL}}\sin(\Omega
t)-\cos(\Omega t)],  \label{ACsol}
\end{equation}
where the characteristic velocity is defined by%
\begin{equation}
V_{1}=\dfrac{\tilde{H}_{z1}\Omega\tau_{\text{CSL}}}{\hbar Q_{0}(1+\alpha
^{2})(1+\tau_{\text{CSL}}^{2}\Omega^{2})}.
\end{equation}
Unlike the case of the transient filed, the oscillational sliding motion is
sustained as a long-term stationary state. This is because in the AC case
the energy associated with the CSL motion is perpetually supplied by the AC
field. It is also seen that the Gilbert damping causes out-of-phase
oscillation[$\cos(\Omega t)$]. For experiment, it may be useful to note%
\begin{equation}
\frac{V_{1}}{V_{0}}=\dfrac{\Omega\tau_{\text{CSL}}}{(1+\tau_{\text{CSL}%
}^{2}\Omega^{2})}\frac{\tilde{H}_{z1}}{\tilde{H}_{z0}}.
\end{equation}
In Fig. \ref{ACCSL}, we show oscillating response of the sliding velocity to
the longitudinal AC field. The transient state rapidly relaxes over the time
scale of $T$, to the stationary forced oscillation with a phase shift due to
the damping.

Here we comment on the relation of the sliding dynamics to the electron spin
resonance (ESR).\cite{KO2009} Provided the whole CSL is in a state of rest,
the longitudinal AC field is able to excite the small amplitude phonon-like
mode (chiral soliton lattice phonon or magnetic kink crystal phonon)
propagating over the CSL. However, because the AC field is uniform, the
resonant phonon absorption occurs only when the momentum absorbed by the
phonon coincides with the reciprocal vector of the super-Brillouin zone of
the CSL. Once the resonance condition is satisfied, the microwave energy
would be consumed to excite the CSL phonons. On the other hand, in the case
of off-resonance, the sliding motion would be driven. We also note that the
excitations associated with the fluctuations of $\varphi$ is totally
irrelevant to the CSL phonon excitation. On the other hand, the sliding
motion is a consequence of the correlated dynamics of coupled $\theta$ and $%
\varphi$.

\section{Numerical analysis of dynamics}

\subsection{Static deformation of CSL}

So far we discussed the CSL dynamics in analytic manner. To justify the
obtained results, it is desirable to perform numerical simulations of the
dynamics. For numerical analysis, we start with the lattice version of Eqs. (%
\ref{EOMtheta}) and (\ref{EOMphi}) written as 
\begin{subequations}
\begin{align}
\dfrac{d\theta_{i}}{d\tau} & =\sqrt{1+\frac{D^{2}}{J^{2}}}\sin\theta
_{i-1}\sin(\varphi_{i}-\varphi_{i-1}+\delta)  \notag \\
& -\sqrt{1+\frac{D^{2}}{J^{2}}}\sin\theta_{i+1}\sin(\varphi_{i+1}-\varphi
_{i}+\delta)  \notag \\
& -\beta_{x}\sin\varphi_{i},  \label{main_system_s_delta1} \\
\dfrac{d\varphi_{i}}{d\tau} & =-(\cos\theta_{i+1}+\cos\theta_{i-1})  \notag
\\
& +\sqrt{1+\frac{D^{2}}{J^{2}}}\cot\theta_{i}\sin\theta_{i-1}\cos(\varphi
_{i}-\varphi_{i-1}+\delta)  \notag \\
& +\sqrt{1+\frac{D^{2}}{J^{2}}}\cot\theta_{i}\sin\theta_{i+1}\cos
(\varphi_{i+1}-\varphi_{i}+\delta)  \notag \\
& -\beta_{x}\cot\theta_{i}\cos\varphi_{i}+\beta_{z},
\label{main_system_s_delta2}
\end{align}
where $\delta=\arctan(D/J)$, $\beta_{x}=\tilde{H}_{x}/JS$, and $\beta _{z}=%
\tilde{H}_{z}/JS$. Here, the time scale $\tau_{0}={\hbar}/{JS}$, and the
dimensionless time $\tau=t/\tau_{0}$ is introduced.

First we consider static spin configurations. In order to perform numerical
computations, we adjust the problem to the form convenient for an iterative
routine, i.e., 
\end{subequations}
\begin{align}
\sin\varphi_{i} & =\left( \mathcal{A}_{i+1}+\mathcal{B}_{i-1}\right)  \notag
\\
& \times\lbrack\left( \mathcal{A}_{i+1}+\mathcal{B}_{i-1}\right) ^{2}+\left( 
\mathcal{C}_{i-1}+\mathcal{D}_{i+1}+\beta_{x}\right) ^{2}]^{-1/2},
\label{NStat1}
\end{align}%
\begin{align}
\cos\varphi_{i} & =\left( \mathcal{C}_{i-1}+\mathcal{D}_{i+1}+\beta
_{x}\right)  \notag \\
& \times\lbrack\left( \mathcal{A}_{i+1}+\mathcal{B}_{i-1}\right) ^{2}+\left( 
\mathcal{C}_{i-1}+\mathcal{D}_{i+1}+\beta_{x}\right) ^{2}]^{-1/2},
\label{NStat2}
\end{align}%
\begin{align}
\cos\theta_{i} & =\left( \cos\theta_{i+1}+\cos\theta_{i-1}+\beta_{z}\right) 
\notag \\
& \times\lbrack\left( \cos\theta_{i+1}+\cos\theta_{i-1}+\beta_{z}\right) ^{2}
\notag \\
& +\left( \mathcal{A}_{i+1}+\mathcal{B}_{i-1}\right) ^{2}+\left( \mathcal{C}%
_{i-1}+\mathcal{D}_{i+1}+\beta_{x}\right) ^{2}]^{-1/2},  \label{NStat3}
\end{align}
where we defined%
\begin{equation}
\left( 
\begin{array}{c}
\mathcal{A}_{i} \\ 
\mathcal{B}_{i} \\ 
\mathcal{C}_{i} \\ 
\mathcal{D}_{i}%
\end{array}
\right) =\sqrt{1+{D^{2}}/{J^{2}}}\sin\theta_{i}\left( 
\begin{array}{c}
\sin\left( \varphi_{i}+\delta\right) \\ 
\sin\left( \varphi_{i}-\delta\right) \\ 
\cos\left( \varphi_{i}-\delta\right) \\ 
\cos\left( \varphi_{i}+\delta\right)%
\end{array}
\right) .
\end{equation}

The spin configuration is found by using the original spin variables and
iteratively re-pointing each along the effective local field due to its
neighbors. Scanning linearly through the chain, spin variable at each site
is updated in sequence, being reset along the net field due partly to some
unchanged neighbors and some that have already been re-pointed. This gives
convergence faster than a synchronized global update.

\begin{figure}[t]
\begin{center}
\includegraphics[width=50mm]{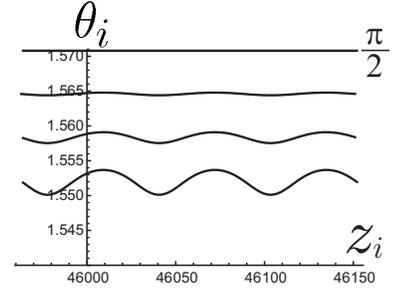}
\end{center}
\caption{Numerical dependencies of $\protect\theta$ on the coordinate $z$.
The length of the chain is $L=10^{5}$ sites. The calculation is performed
with the same parameters as in Fig. \protect\ref{TetReal}. The perpendicular
field, $H_{x}/H_{c}=0;\,0.1;\,0.2;\,0.3$ (from above to below), is
normalized to the critical field $H_{c}$.}
\label{StaticTheta}
\end{figure}

The most difficult computational problem in carrying out this program is to
find the initial configuration that relaxes to a target spin configuration.
It is meaningful to impose appropriate boundary conditions too. Obviously
this is a rich problem with a wide choice of options. In our simulations we
choose a starting configuration as the simple spiral, $%
\varphi_{i}=q(i-z_{0}) $ and $\theta_{i}=\pi/2$, and take the free boundary
conditions. The coordinate $z_{0}$ corresponds to a position with $%
\varphi_{i}=\pi$ in a middle of the chain of the length $L$. The value of $%
q=0.1$ is taken throughout the numerical calculations to make more
appreciable a spatial modulation of solutions.

\begin{figure}[t]
\begin{center}
\includegraphics[width=45mm]{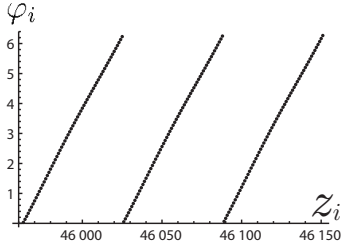}
\end{center}
\caption{Coordinate dependence of the $\protect\varphi$ (shown as mod $2%
\protect\pi$) obtained numerically for the chain of length $L=10^{5}$ sites
at the same parameters as in Fig. \protect\ref{TetReal}, and $%
H_{x}/H_{c}=0.1 $. From the almost strict linear behavior, one sees that $%
\protect\varphi$ acquires almost no change.}
\label{StaticPhi}
\end{figure}

We revealed that a convergence of the iteration process is very slow. The
iterations stop if the sum 
\begin{equation}
\sigma=\sqrt{\sum_{i=1}^{L}\left(
\varphi_{i}^{(k)}-\varphi_{i}^{(k-1)}\right) ^{2}+\sum_{i=1}^{L}\left(
\theta_{i}^{(k)}-\theta _{i}^{(k-1)}\right) ^{2}}
\end{equation}
taken over the chain on the $k$-th step is less than tolerance $10^{-8}$. To
reach the accuracy, around $205\times10^{6}$ iterations are required.

\begin{figure}[t]
\begin{center}
\includegraphics[width=50mm]{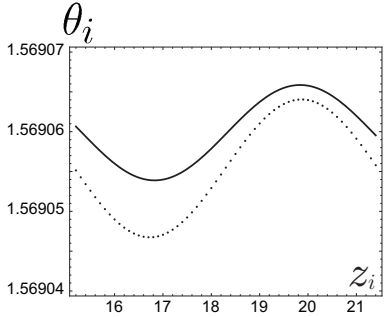}
\end{center}
\caption{A comparison between numerical data for the static case (dotted
line) and analytical expression given by Eq.(\protect\ref{EllipticTheta})
(solid line). The fields are $\protect\beta_{x}=b\cos\protect\delta_{b}$ and 
$\protect\beta_{z}=b\sin\protect\delta_{b}$, $b=10^{-4}$, $\protect\delta%
_{b}=\protect\pi/18$.}
\label{Static3Plots}
\end{figure}

The numerical behavior of $\theta$ shown in Fig. \ref{StaticTheta}
reproduces the theoretical finding (see Fig.~\ref{TetReal}), i.e. an
increasing of the longitudinal field enhances a modulation of a conical
structure along the $z$ axis. The calculation confirms another assumption of
the Sec. IV, namely, the longitudinal field causes no changes in the
variable $\varphi$ (Fig.~\ref{StaticPhi}). The numerical data are imposed on
the theoretical predictions [Eq. (\ref{EllipticTheta})] as shown in Fig.~\ref%
{Static3Plots}. Evidently, they reveal a good agreement with each other
(Fig. \ref{Static3Plots}). 
\begin{figure}[b]
\begin{center}
\includegraphics[width=85mm]{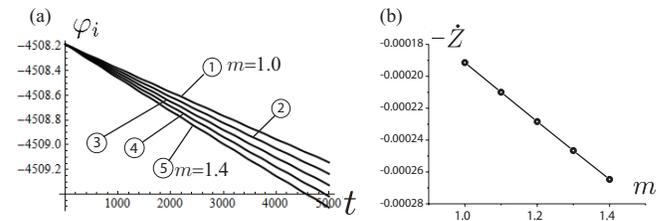}
\end{center}
\caption{(a) Linear time dependence of $\protect\varphi(t)$ for the central
site of a chain of length $L~=~1000001$ obtained by numerical simulations.
The fields are $\protect\beta_{x}=b\cos\protect\delta_{b}$ and $\protect\beta%
_{z}= m b\sin\protect\delta_{b}$, $b=10^{-3}$, $\protect\delta_{b}=\protect%
\pi/18$. With a growth of $\protect\beta_{z}$ (or $m$) a slope of the curves
increases linearly according to analytical result. The lines correspond to $%
m=1.0-1.4$ with the step 0.1 from above to below, respectively. (b) The
dependence $\dot{Z}$ on the factor $m$ extracted from the $\protect\varphi%
(t) $ data.}
\label{ACsolrev}
\end{figure}

\subsection{Dynamics}

A search for dynamical solutions is carried out by using the eighth-order
Dormand-Prince method implemented in Ref. \cite{Press2007}. The embedded
Runge-Kutta integrator with an adaptive step-size control ensures a relative
tolerance $10^{-12}$. The integration spans a period of time from zero till $%
2\times10^{4}\,\tau_{0}$. The length of chains used in the computations
amounts to $10^{5}+1$ sites. A time evolution of magnetization was monitored
by recording the time dependencies of the $\theta$ and $\varphi$ variables
for the central site.

\begin{figure}[t]
\begin{center}
\includegraphics[width=85mm]{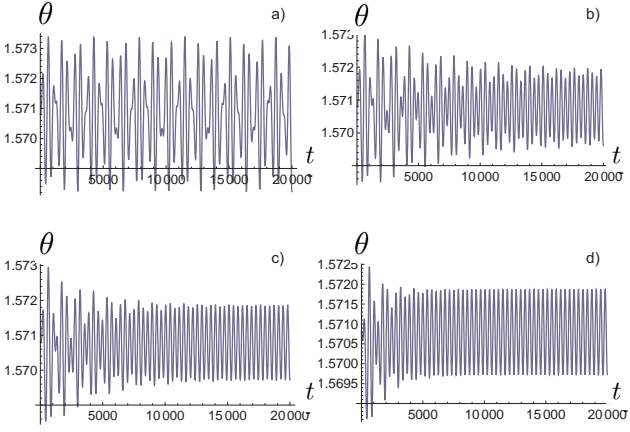}
\end{center}
\caption{The time-dependent variation of the $\protect\theta$ for different
damping parameters: (a) $\protect\alpha=0$, (b) $\protect\alpha=0.01$, (c) $%
\protect\alpha=0.02$, (d) $\protect\alpha=0.05$. The fields are $\protect%
\beta_{x}=b\cos\protect\delta_{b}$ and $\protect\beta _{z}=-b\sin\protect%
\delta_{b}$, $b=10^{-3}$, $\protect\delta_{b}=\protect\pi/18$, the ratio $%
\protect\beta _{z}/\Omega$ equals to 0.01.}
\label{AlphaEvolutTheta}
\end{figure}

In Fig. \ref{ACsolrev}, we show numerical result for the velocity. It is
seen that the velocity decreased linearly with increasing $\beta _{z}$\
field. In the calculations $\beta _{x}$ was held constant.

\begin{figure}[t]
\begin{center}
\includegraphics[width=85mm]{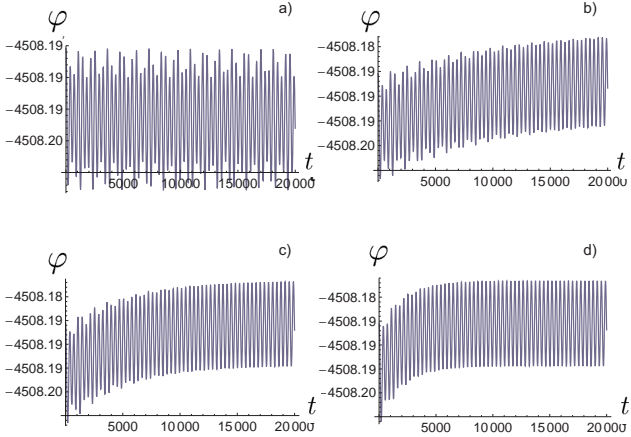}
\end{center}
\caption{The time-dependent variation of the intrinsic mode of $\protect%
\varphi $ for different damping parameters: (a) $\protect\alpha =0$, (b) $%
\protect\alpha =0.01$, (c) $\protect\alpha =0.02$, (d) $\protect\alpha =0.05$%
. The fields and frequency are the same as in the previous Figure.}
\label{AlphaEvolutPhi}
\end{figure}

Figs. \ref{AlphaEvolutTheta}, \ref{AlphaEvolutPhi} show the time dependence
of $\theta $ and $\varphi $, respectively, under the oscillating $\beta _{z}$%
\ for different values of the damping parameter $\alpha $. These
calculations make evident a salient feature of the forced oscillations. In
the initial stage of time evolution, the longitudinal magnetic field excites
intrinsic eigenmodes that are superimposed on the field-driven oscillations.
The eigenmodes fade away in the steady-state regime and they damped more
rapidly the greater the parameter $\alpha $. The period of the forced
oscillations $2\pi /\Omega $ exactly corresponds the period of the driving
force. In appendix B, we present the detailed analysis of this forced
osciilation in line with the numerical analysis. 

\section{Spin motive force}

Now that we have obtained the CSL\ dynamics under the crossed magnetic
field, we will go on to discuss possible SMF generation. Because the sliding
motion of the CSL accompanies the dynamical deformation of the spin texture,
we naturally expect the SMF to occur in the configuration presented in Fig. %
\ref{CSL}. Generally speaking, when conduction electrons adiabatically see a
spatially modulated spin structure along the $z$ axis, spinor wave function
locally follows the background. Consequently, the spinor space turns out to
be curved. The corresponding curvature is represented by the gauge (Berry)
connections. In the adiabatic picture, the Berry curvature in the spinor
space acts as an effective electric field,\cite{Volovik1987,Xiao2010} 
\begin{equation}
E_{\sigma }(z,t)=-\frac{\hbar \sigma }{2e}\Omega _{zt}=\frac{\hbar \sigma }{%
2e}\sin \theta \left( \partial _{z}\theta \,\partial _{t}\varphi -\partial
_{z}\varphi \,\partial _{t}\theta \right) .  \label{ElectroViaBerry}
\end{equation}%
Then, we obtain a general expression for the SMF is given by 
\begin{equation}
\varepsilon _{\sigma }(t)=\int_{0}^{L}dz\,E_{\sigma }(z,t)=\frac{\hbar
\sigma }{2e}\frac{d}{dt}\left( \int_{\Gamma }\cos \theta d\varphi \right) ,
\label{SMF}
\end{equation}%
where the contour $\Gamma $ is taken on the sphere presenting a space of the
order parameter $\boldsymbol{n}$. The voltage is related via the Stokes
theorem with a change of area (Berry cap) $\mathcal{S}$ on the sphere
enclosed by the contour,\cite{Yang} 
\begin{equation}
\varepsilon _{\sigma }(t)=-\frac{\hbar \sigma }{2e}\frac{d\mathcal{S}}{dt}.
\label{Faraday}
\end{equation}%
This involves an analogue of Faraday's law for the emergent electromagnetic
field, where a magnetic field of a Dirac monopole with a charge $\hbar /2$
plays a role of the flux enclosed by the Berry cap $\mathcal{S}$.

\begin{figure}[t]
\begin{center}
\includegraphics[width=75mm]{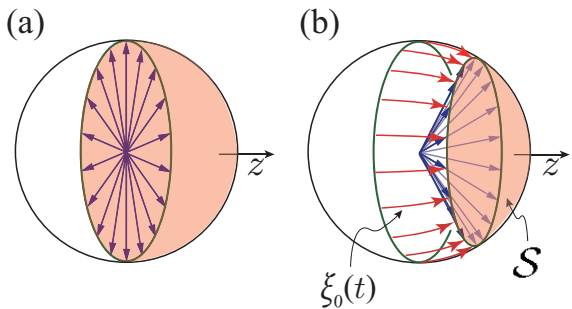}
\end{center}
\caption{Switching of the time-dependent longitudinal field $H_{z}(t)$
causes the change in Berry cap $\mathcal{S}$ from (a) to (b) in a
time-dependent manner. This time-dependence causes the SMF along the chiral
axis. Each allow represents local spin configuration $\boldsymbol{n}(z)$.
The Berry cap is associated with the area traced out by $\boldsymbol{n}(z)$. 
}
\label{Berry}
\end{figure}

In the present case of the CSL\ dynamics, by using the collective
representation [Eqs. (\ref{collective_phi}) and (\ref{collective_theta})],
the SMF for majority spins (the case of minority spins has the opposite
sign) is computed as%
\begin{equation}
\varepsilon(t)\simeq-\frac{\hbar}{2e}Q_{0}\dot{\xi}_{0}(t)%
\int_{0}^{L}u_{0}(z)dz,  \label{SMFgeneral}
\end{equation}
This expression indicates that \textit{the SMF arises only for} $\dot{\xi}%
_{0}\neq0$, i.e., the time-dependence of the Berry cap is essential to cause
the SMF [see Fig.\ref{Berry}]. This observation is consistent with the
discussion given in Ref.\cite{Yang}.

Using Eq. (\ref{zero_mode}), we obtain 
\begin{equation}
\int_{0}^{L}u_{0}(z)dz=\frac{4}{Q_{0}}\sqrt{\frac{KE}{L}}\mathcal{Q}\text{,}
\end{equation}
where 
\begin{equation}
\mathcal{Q}=\frac{\varphi_{0}(L)-\varphi_{0}(0)}{2\pi},
\end{equation}
\ is the topological charge\ representing the number of solitons over the
whole length of a sample. Using the relation, $\dot{\xi}_{0}=-\alpha Q_{0}%
\sqrt{L}\dot{Z}$, obtained from Eq. (\ref{CEOM1}), Eq. (\ref{SMFgeneral})
finally reduces to%
\begin{equation}
\varepsilon(t)\simeq\mathcal{Q}\frac{\hbar}{e}\pi\alpha Q_{0}\dot{Z}(t),
\label{SMF formula}
\end{equation}
where we used the relation $\sqrt{KE}\simeq\pi/2$ in the case of a weak
transverse field. As expected, the SMF is directly proportional to the
macroscopic number of soliton, $\mathcal{Q}$. It is worth whole to compare
the obtained formula [Eq. (\ref{SMF formula})] with the one used for the SMF
induced by domain wall motion.\cite{Yang} In the present case of CSL, the
SMF to be strongly amplified by the prefactor $\mathcal{Q}$. Furthermore, it
should be stressed that \textit{the dissipative dynamics is essential to
drive SMF.} Actually, the SMF is proportional to the Gilbert damping
parameter, $\alpha$. In appendix C, we show that the dissipationless rigid
motion of the CSL never produces SMF.

In the case of the time-dependent longitudinal field, $%
H_{z}(t)=H_{z0}(1-e^{-t/T}),$ plugging Eq. (\ref{Velocity}) into Eq. (\ref%
{SMF formula}), we immediately obtain%
\begin{equation}
\varepsilon (t)\simeq -\varepsilon _{0}\frac{\tau _{\text{CSL}}}{\tau _{%
\text{CSL}}-T}\left( e^{-t/\tau _{\text{CSL}}}-e^{-t/T}\right) ,
\end{equation}%
where%
\begin{equation}
\varepsilon _{0}=\alpha \mathcal{Q}\frac{\hbar }{e}\pi Q_{0}V_{0}.
\label{SMFamp}
\end{equation}

In Cr${}_{1/3}$NbS$_{2}$, using $Q_{0}\simeq 1.3\times 10^{8}$[m$^{-1}$] and 
$V_{0}\simeq $0.13[m$\cdot $s$^{-1}\cdot $Oe$^{-1}$] as estimated in Sec.
III.B and assuming $\alpha \simeq 10^{-2}$, we have an estimation, $%
\varepsilon _{0}$ $\simeq 0.36\mathcal{Q}H_{z0}$ [nV] when $H_{z0}$\ is
measured in Oersted. In the case where the sample size is $L\simeq $1mm, the
upper bound of $\mathcal{Q}$\ along the helical axis amounts to $L/L{{_{%
\text{CHM}}}}\simeq 10^{5}$. Therefore, we expect that $\varepsilon _{0}$
amounts to $1$[mV] for $H_{z0}\simeq 10^{2}$Oe as an example. To
experimentally sustain the SMF, it may be desirable to apply a sequence of
the pulse fields.

Here we comment on the physical reason why the SMF is proportional to the
Gilbert damping factor $\alpha $ in Eq. (\ref{SMFamp}). An essential point
is that emergence of the SMF\ is a direct consequence of time-varying Berry
cap,\cite{Yang} which needs finite $\xi _{0}$. Now, as is clearly seen from
basic EOMs, (\ref{CEOM1}) and (\ref{CEOM2}), if the Gilbert damping were
absent, $\xi _{0}$ and $Z$ are dynamically decoupled and consequently $\xi
_{0}=\xi _{0}(0)=0$ for all the time, i.e., the sliding motion can never be
sustained. This situation is totally different from the case of 180-degree
Bloch wall. The CSL\ is regarded as an array of 360-degree walls and we need
some mechanism which enables the magnetic moments to rotate around the
chiral axis. Only one possible mechanism to realize this rotation is the
Gilbert damping process. This is the reason why the SMF is proportional to $%
\alpha $. More intuitively speaking, at the first stage the longitudinal
field $H_{z}$ directly couples to $\xi _{0}$ and causes the out-of-plane
canting of the magnetic moments [see the Lagrangian (\ref{Collective
Lagrangian})]. At the second stage, because of the Gilbert damping, the
magnetic moments start damped precession to relax back into their original
directions. This motion triggers the moments to rotate around the chiral
axis and eventually leads to the collective sliding.

\section{Concluding remarks}

In this paper, we demonstrated that the chiral soliton lattice (CSL)
exhibits coherent sliding motion simply applying a \textit{time-dependent}
magnetic longitudinal field, in addition to a \textit{static} transverse
field. The driving force of the sliding is given by the Zeeman coupling of
the collective coordinate $\xi _{0}$ with the longitudinal field. This
mechanism is intuitively understood by DBK mechanism of the moving domain
wall (DW).\cite{Doring48,Becker50,Kittel50} In the DBK mechanism, once the
domain wall begins to move, so-called demagnetization field is dynamically
generated inside the wall. The demagnetization field supplies spin torque to
keep the inertial motion. Actually, the DBK mechanism of a single DW was
analyzed by exactly the same procedure presented in this paper.\cite{KO2010}
In the CSL dynamics, the longitudinal magnetic field kicks off the
demagnetization and drives coherent sliding motion. To demonstrate the
coherent sliding motion, we first used the collective coordinate method, and
then confirmed the result by computational analysis.

The time duration of the sliding motion is characterized by the intrinsic
relaxation time determined by Eq. (\ref{taugap}). Providing the gap energy $%
\varepsilon_{0}^{(\theta)}$ varies between $0.1\sim10$K and the Gilbert
damping constant varies between $10^{-4}\sim10^{-2}$, we expect the time
duration varies between $10^{-9}\sim10^{-6}$ sec. To realize longer-lasting
sliding motion, a smaller value of $\alpha$ and a smaller gap frequency $%
\omega_{\text{gap}}$ may be desirable. This estimation may give a guiding
principle for materials synthesis. Sequential pluses of the longitudinal
magnetic fields may remedy a quick decay of the sliding.

The sliding motion may be signaled by the spin-density accumulation inside
each soliton (kink) and emergence of periodic arrays of the induced magnetic
dipoles carrying the transport spin current.\cite{BKO2008} From theoretical
viewpoints, the coherent sliding is always possible to occur as a direct
consequence of the phase rigidity and Galilean symmetry in any type of
density waves, including spin/charge density waves and even inhomogeneous
superconducting states. However, in many types of such systems, the sliding
motion does not transport experimentally measured quantity.\cite{BasicNotion}
In this respect, it is remarkable that the coherent sliding of the CSL
accompanies the dynamically generated magnetization.

Another observable consequence of the sliding is appearance of the SMF along
the helical axis. We showed that the time-dependent sliding velocity $\dot {Z%
}(t)$ causes time-varying Berry cap which causes the SMF. We stressed that
the dissipative dynamics plays an essential role to drive SMF. A salient
feature of the CSL is appearance of the strongly amplified SMF which is
directly proportional to the macroscopic number of soliton. Consequently,
the SMF is expected to reach the order of mV. As reported in Ref.\cite%
{Togawa2012}, CHM and CSL are quite robust against structural dislocation
and crystal defects. Its high stability and robustness are direct
manifestation of the macroscopic order of spin magnetic moments in CHM and
CSL state. We hope that the present proposal may lead to spintronics
application based on chiral magnetic crystals.

\begin{acknowledgments}
J.~K. acknowledges Grant-in-Aid for Scientific Research (A) (No.~22245023)
and Grant-in-Aid for Scientific Research on Innovative Areas (No.~24108506)
from the Ministry of Education, Culture, Sports, Science and Technology,
Japan. Vl.~E.~S. acknowledges Grant RFBR No.~12-02-31565 mol\_a. We
acknowledge helpful discussions with Y. Togawa and J. Akimitsu.
\end{acknowledgments}

\appendix

\section{Static deformation}

Plugging the expressions, $\theta(z)=\pi/2+s\tilde{\theta}(z)$\ and $%
\varphi(z)=\varphi_{0}(z)+s\tilde{\varphi}(z)$, into the static counterparts
of Eqs. (\ref{EOMtheta}) and (\ref{EOMphi}), and retaining the first order
corrections with respect to $s$, we have 
\begin{subequations}
\begin{gather}
\partial_{\bar{z}}^{2}\tilde{\varphi}=\left( 2\kappa^{2}\text{sn}^{2}{\bar {z%
}}-\kappa^{2}\right) \tilde{\varphi}  \label{Lame_phi} \\
\partial_{\bar{z}}^{2}\tilde{\theta}=\left( 2\kappa^{2}\text{sn}^{2}{\bar{z}}%
-\kappa^{2}\right) \tilde{\theta}-\frac{\kappa^{2}}{\beta_{x}}\left( {Q}%
_{0}^{2}\theta_{1}-\beta_{z}\right) ,  \label{Lame_theta}
\end{gather}
where the dimensionless variables $\bar{z}=\sqrt{\beta_{x}}z/\kappa$, $%
\beta_{z}=\tilde{H}_{z}/JS,$ $\beta_{x}=\tilde{H}_{x}/JS$ are introduced.
Eq. (\ref{Lame_phi}) is the homogeneous Lam\'{e} equation, while Eq. (\ref%
{Lame_theta}) is a non-homogeneous Lam\'{e} equation. The solution of Lam%
\'{e} equation is well known to be given in a form,\cite{WW} 
\end{subequations}
\begin{equation}
\tilde{\varphi}_{1,2}({\bar{z}})=\frac{\mathrm{H}({\bar{z}}\pm a)}{\Theta({%
\bar{z}})}e^{\mp{\bar{z}}Z(a)},  \label{Fundament}
\end{equation}
where $\mathrm{H}$ and $Z$ are Jacobi's eta and zeta functions, respectively
with the parameter $a$ being determined by dn$^{2}a=-16E^{2}/\pi^{2}.$

A solution for the non-homogeneous equation (\ref{Lame_theta}) is obtained
by using homogeneous solutions $\tilde{\varphi}_{1,2}$. In the inhomogeneous
term, we ignore ${Q}_{0}^{2}\theta_{1}$ as compared with $\beta_{z}$. This
treatment is justified because $\beta_{z}$ and $\theta_{1}$ are of the same
order, and ${Q}_{0}\ll1$. Using a method of variation of parameters for
non-homogeneous second-order differential equation, we readily construct the
solution as%
\begin{equation}
\tilde{\theta}({\bar{z}})=\dfrac{\beta_{z}}{\beta_{x}}\kappa^{2}W^{-1}\left( 
\tilde{\varphi}_{2}({\bar{z}})\int^{{\bar{z}}}d{\bar{z}}\,\tilde{\varphi}%
_{1}({\bar{z}})-\tilde{\varphi}_{1}({\bar{z}})\int^{{\bar{z}}}d{\bar{z}}\,%
\tilde{\varphi}_{2}({\bar{z}})\right) ,  \label{thetaint}
\end{equation}
where $W$ is the Wronskian,%
\begin{equation}
W=\tilde{\varphi}_{1}({\bar{z}})\tilde{\varphi}_{2}^{^{\prime}}({\bar{z}})-%
\tilde{\varphi}_{2}({\bar{z}})\tilde{\varphi}_{1}^{^{\prime}}({\bar{z}}).
\label{DefWron}
\end{equation}
The Lam\'{e} equation guarantees $dW/d{\bar{z}}=0,$i.e. the Wronskian is
independent of ${\bar{z}}$ and therefore $W=W(0)$. By plugging the
expressions%
\begin{equation}
\tilde{\varphi}_{1}(0)=-{\tilde{\varphi}}_{2}(0)=\dfrac{H(a)}{\Theta (0)}=%
\dfrac{\theta_{1}\left( \frac{\pi a}{2K}\right) }{\theta_{4}(0)},
\end{equation}
and%
\begin{equation}
{\tilde{\varphi}}_{1}^{^{\prime}}(0)={\tilde{\varphi}}_{2}^{^{\prime}}(0)=%
\frac{\pi}{2K}\frac{H^{^{\prime}}(a)}{\Theta(0)}-\frac{H(a)}{\Theta (0)}Z(a),
\end{equation}
into Eq.(\ref{DefWron}), we finally obtain 
\begin{equation}
W=\frac{2}{\vartheta_{4}^{2}}\theta_{1}^{2}\left( \frac{\pi a}{2K}\right) %
\left[ \frac{\pi}{2K}\frac{\theta_{1}^{^{\prime}}\left( \frac{\pi a}{2K}%
\right) }{\theta_{1}\left( \frac{\pi a}{2K}\right) }-Z(a)\right] ,
\label{WronLame}
\end{equation}
where $\theta_{i}(x)$ ($i=1,2,3,4$) denote the elliptic Theta functions and $%
\vartheta_{4}\equiv\theta_{4}(0)=\sqrt{2\kappa^{^{\prime}}K/\pi}$. Here $%
\kappa^{^{\prime}}$ is the complementary modulus.

Final task is to perform integrals in (\ref{thetaint}) is performed by using
the Fourier transformation of $\tilde{\varphi}_{1,2}({\bar{z}})$ to give Eq.
(\ref{EllipticTheta}). The derivation is similar to the calculation of
Fourier coefficients for the Jacobi $\text{sn}$ function.\cite{WW} We start
with the Fourier transformation 
\begin{equation}
\tilde{\varphi}_{1,2}({\bar{z}})=\sum_{n=-\infty}^{+\infty}c_{1,2\,n}e^{\bar{%
z}\left( \frac{i\pi n}{2K}\mp Z(a)\right) }.  \label{FTSeries}
\end{equation}
By definition, we have 
\begin{equation*}
2\pi c_{-n}=\int_{-\pi}^{\pi}\frac{\theta_{1}(x+a)}{\theta_{4}(x)}%
e^{inx}\,dx.
\end{equation*}
To evaluate the integral, we use the contour taken as a parallelogram with
the corner points $-\pi$, $\pi$, $\pi+\pi\tau$ and $-2\pi+\pi\tau$, where $%
\tau=iK^{^{\prime}}/K$. The singular points inside the contour are $%
z_{1}=-\pi+\pi\tau/2$ and $z_{2}=\pi\tau/2$. After straightforward
computations, we obtain 
\begin{align}
c_{1,2\,n} & =0\text{ for even }n  \label{FTcoef} \\
c_{1,2\,n} & =-i\frac{\theta_{4}\left( \frac{\pi a}{2K}\right) }{%
\theta_{1}^{^{\prime}}\sinh\left[ \pi\left( nK^{^{\prime}}\mp ia\right) /2K%
\right] }\text{ for odd }n.
\end{align}
for odd $n$. Finally we have%
\begin{align}
C_{1} & =-\sum_{n=-\infty}^{+\infty}c_{2n}\,\dfrac{e^{\bar{z}\left( \frac{%
i\pi n}{2K}+Z(a)\right) }}{\frac{i\pi n}{2K}+Z(a)},  \label{FTC1} \\
C_{2} & =\sum_{n=-\infty}^{+\infty}c_{1n}\,\dfrac{e^{\bar{z}\left( \frac{%
i\pi n}{2K}-Z(a)\right) }}{\frac{i\pi n}{2K}-Z(a)},  \label{FTC2}
\end{align}
where in the summation only terms with odd $n$\ are retained.

\section{AC field driven oscillations of the CSL}

We derive a solution of the LLG equations (\ref{EOMtheta},\ref{EOMphi}) for
the periodic longitudinal magnetic field $\beta_z (\tau)= \beta_{z0} \sin
\Omega \tau$. The solution is sought in the form $\theta=\pi/2 +\theta_1$
and $\varphi= \varphi_0+\varphi_1$, where the additions $\theta_1$, $%
\varphi_1$ are of the same order of magnitude as the magnetic field $\beta_z$%
. To provide an analytical treatment, we consider the limit of the small $%
\beta_x$ when the approximations $\varphi_0(z) \approx Q_0 z$ and $\theta_0
= \pi/2$ are relevant. Moreover, we assume a smallness of the Gilbert
damping, when the problem becomes iterative. At the first stage, we find
solutions for $\theta_1$, $\varphi_1$ at $\alpha=0$ and plug them into Eqs.(%
\ref{EOMtheta},\ref{EOMphi}) to obatin new values valid for non-zero $\alpha$%
.

At $\alpha=0$ Eqs.(\ref{EOMtheta},\ref{EOMphi}) read as 
\begin{equation}  \label{EqThetT1}
\frac{\partial \theta_1}{\partial \tau} = - \frac{\partial^2 \varphi_1}{%
\partial z^2} - \beta_x \cos Q_0 z \, \varphi_1,
\end{equation}
\begin{equation}  \label{EqPhiT1}
\frac{\partial \varphi_1}{\partial \tau} = -Q^2_0 \theta_1 + \frac{%
\partial^2 \theta_1}{\partial z^2} + \beta_x \cos Q_0 z \, \theta_1 + \beta_z
\end{equation}
and can be easily resolved through the substitutions 
\begin{equation*}
\varphi^{(0)}_1(z,\tau) = \left(A_1 + A_2 \beta_x \cos Q_0 z \right) \cos
\Omega \tau,
\end{equation*}
\begin{equation}  \label{Separat}
\theta^{(0)}_1(z,\tau) = \left(B_1 + B_2 \beta_x \cos Q_0 z \right) \sin
\Omega \tau,
\end{equation}
where $A_{1,2}$ and $B_{1,2}$ are the unknowns.

This results straightforwardly in 
\begin{equation}  \label{CT1}
\varphi^{(0)}_1 (z,\tau) = - \left( \frac{\beta_{z0}}{\Omega} + \frac{%
\beta_{z0}}{\Omega} \frac{2 \beta_x Q^2_0}{(2 Q^4_0 - \Omega^2)} \cos Q_0 z
\right) \cos \Omega \tau,
\end{equation}
\begin{equation}  \label{CT2}
\theta^{(0)}_1 (z,\tau) = - \frac{\beta_x \beta_{z0}}{2 Q^4_0-\Omega^2} \cos
Q_0 z \, \sin \Omega \tau.
\end{equation}
A requirement of smallness of the corrections amounts to $\beta_{z0} \ll
\Omega$, $2 \beta_x \beta_{z0} Q^2_0 \ll \Omega (2 Q^4_0 - \Omega^2)$, and $%
\beta_x \beta_{z0} \ll 2 Q^4_0 - \Omega^2$. Taking $Q_0 \sim 10^{-2}$ and $%
\Omega \sim 10^{-4}$ in dimensionless units (or 1 GHz in physical units $%
\Omega \tau_0$), we can suppose, for example, $\beta_{z0}/\Omega \sim 0.1$, $%
\beta_x \sim 10^{-4}$ (100~Oe), $\beta_{z0} \sim 10^{-5}$ (10~Oe).

\begin{figure}[t]
\begin{center}
\includegraphics[width=65mm]{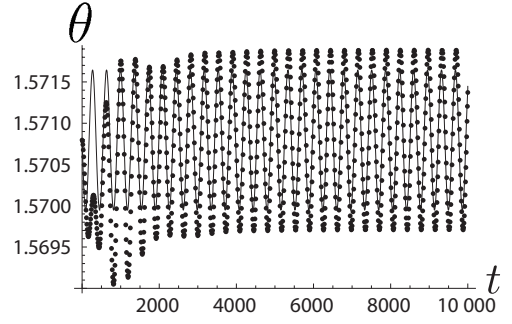}
\end{center}
\caption{The time-dependent variation of the polar angle for the central
site ($N=50000$): numerical data (dots) and analytical result (line) given
by Eq.(\protect\ref{SinOmegaDampTheta}). The fields are taken as in Fig.8 in
the main text, $\protect\alpha=0.1$.}
\label{ThetaAlpha0}
\end{figure}

\begin{figure}[t]
\begin{center}
\includegraphics[width=65mm]{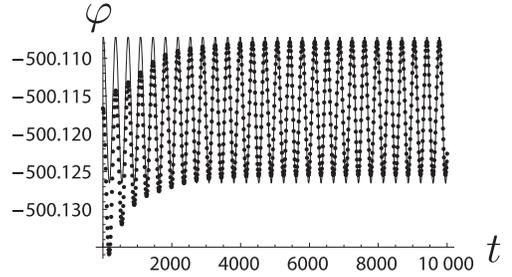}
\end{center}
\caption{The time-dependent variation of the azimuthal angle for the central
site ($N=50000$): numerical data (dots) and analytical result (line) given
by Eq.(\protect\ref{SinOmegaDampPhi}). The parameters are the same as in the
previous Figure.}
\label{PhiAlpha0}
\end{figure}

By assuming smallness of the Gilbert parameter $\alpha$, we organize the
iterative procedure to find solutions of the system (\ref{EOMtheta},\ref%
{EOMphi}) with the time derivatives in the right-hand sides estimated from
Eqs. (\ref{CT1},\ref{CT2}). The calculation yields 
\begin{equation}  \label{SinOmegaDampTheta}
\theta_1 (z,\tau) = \theta^{(0)}_1 (z,\tau) - \alpha \frac{\beta_{z0}}{\Omega%
} \cos \Omega \tau
\end{equation}
\begin{equation*}
\times \left[1 - \frac{\beta_x Q^2_0}{\Omega^2-2 Q^4_0} \left( 2 + \frac{%
3\Omega^2}{\Omega^2-2 Q^4_0} \right) \cos Q_0 z \right] ,
\end{equation*}
\begin{equation}  \label{SinOmegaDampPhi}
\varphi_1 (z,\tau) = \varphi^{(0)}_1 (z,\tau) + \alpha \frac{\beta_{z0}}{%
\Omega} \sin \Omega \tau
\end{equation}
\begin{equation*}
\times \left[ \frac{Q^2_0}{\Omega} - \frac{\beta_x \Omega}{\Omega^2-2 Q^4_0}
\left( 1 + 2 \frac{Q^4_0}{\Omega^2} + \frac{\Omega^2+4 Q^4_0}{\Omega^2-2
Q^4_0} \right) \cos Q_0 z \right] .
\end{equation*}
A direct comparison between the numerical results and the analytical
predictions is done in Figs. \ref{ThetaAlpha0}, \ref{PhiAlpha0}. Obviously,
there is a good agreement in the steady-state regime, when eigenmodes fade
away. 

\section{Absence of SMF for rigid motion}

We here note that the dissipationless rigid motion of the CSL never causes
SMF. The results can be obtained from general considerations. In Villain's
representation,\cite{Villain} the spin component $S^{z}=\hbar S\cos\theta$
and the angle $\varphi$ made by the projection of the spin in the $(x,y)$
plane are conjugated canonical variables. Eqs. (\ref{EOMtheta}) and (\ref%
{EOMphi}) are written in the new variables acquire the Hamiltonian form 
\begin{equation}
\frac{\partial\varphi}{\partial t}=\frac{\partial\mathcal{H}}{\partial S^{z}}%
,\quad\frac{\partial S^{z}}{\partial t}=-\frac{\partial\mathcal{H}}{%
\partial\varphi},  \label{HamiltonianForm}
\end{equation}
whereas the fictitious electric field (\ref{ElectroViaBerry}) is presented
as follows 
\begin{equation}
E(z,t)=\frac{1}{2S}\left( \frac{\partial S^{z}}{\partial t}\frac {%
\partial\varphi}{\partial z}-\frac{\partial S^{z}}{\partial z}\frac {%
\partial\varphi}{\partial t}\right) .
\end{equation}
The spin motive force generated along the path of length $L$ reduces to the
contour integral in the phase space of the conjugated variables $(\varphi
,S^{z})$ 
\begin{align}
\varepsilon(t) & =\int_{0}^{L}dz\,E(z,t)=\frac{1}{2S}\oint_{\Gamma}\left( 
\frac{\partial S^{z}}{\partial t}d\varphi-\frac{\partial\varphi}{\partial t}%
dS^{z}\right)  \notag \\
& =-\frac{1}{2S}\int_{\mathcal{S}}\left[ \frac{\partial}{\partial\varphi }%
\left( \frac{\partial\varphi}{\partial t}\right) +\frac{\partial}{\partial
S^{z}}\left( \frac{\partial S^{z}}{\partial t}\right) \right] d\varphi
dS^{z}.
\end{align}
Plugging Eqs.(\ref{HamiltonianForm}) into the formula we obtain $\varepsilon
(t)=0$. This rigorous result shows dissipationless Hamiltonian dynamics of
any spin texture never causes finite SMF.

\end{document}